\documentclass[doublecol]{epl2} 

\usepackage{graphicx}
\usepackage{bm}
\usepackage{amsfonts}
\usepackage{amsmath}
\usepackage[colorlinks=true,citecolor=blue]{hyperref}
\hypersetup{colorlinks=true,citecolor=blue,linkcolor=red,urlcolor=blue}

\title{Gate controlled supercurrent reversal in MoS$_2$-based Josephson junctions}
\shorttitle{Gate controlled supercurrent reversal in MoS$_2$} 

\author{Babak Zare Rameshti\inst{1} \and  Ali G. Moghaddam\inst{1} \and Malek Zareyan\inst{1}}
\shortauthor{B. Zare Rameshti \etal}

\institute{                    
  \inst{1} Department of Physics, Institute for Advanced Studies in Basic Sciences (IASBS), Zanjan 45137-66731, Iran}

\abstract{Motivated by recent experiments revealing superconductivity in MoS$_2$, we investigate the Josephson effect in the monolayer MoS$_2$ at the presence of an exchange splitting. We show that the supercurrent reversal known as $0-\pi$ transition can occur by varying the doping via gate voltages. This is in contrast to common superconductor/ferromagnet/superconductor junctions in which successive $0-\pi$ transition take place with the variation of junction length or temperature. In fact for the case of MoS$_2$ we find that both the amplitude and the period of oscillations show a dependence on the doping which explains the predicted doping induced supercurrent reversal. These effects comes from the dependence of density and Fermi velocity on the doping strength beside the intrinsic spin splitting in the valence band which originates from spin-orbit interaction. }

\begin{document}

\maketitle

Exotic and peculiar behaviors in low and specially two dimensional (2D) systems have been in the heart of interests of condensed matter physicists for decades \cite{xu13}. But by isolation of one- and few-atomic layer thick materials at 2004, many new and potentially of great technological importance
phenomena have been revealed in 2D systems \cite{neto09}. 
Recently few- and monolayers of transition metal dichalcogenides (TMD) are synthesized \cite{mak10,wang10,wang12} and it has been observed that they can have very diverse properties including metallic, semiconducting and even superconducting behaviors \cite{wang12, chhowalla}. Due to its promising application in optoelectronics as a direct band gap semiconductor, monolayer Molybdenum Disulfide (ML-MoS$_2$) has attracted more interest among monolayers of other TMD \cite{mak12,mak13,zeng12,cao12,wu13}. Moreover, it has a strong intrinsic spin-orbit coupling which results in the spin-split valence subbands separated by $\sim 160 $ meV from each other. Combined with the fact that we have two valleys $K$ and $K'$ in the band structure, a rich spin and valley physics including valley and spin Hall effects is exhibited in ML-MoS$_2$ devices \cite{xiao12}. In addition it can have promising applications in nano-electronics \cite{radisavljevic}, spintronics \cite{loss13}.
\par
\begin{figure}
\begin{center}
\includegraphics[width=0.8\linewidth]{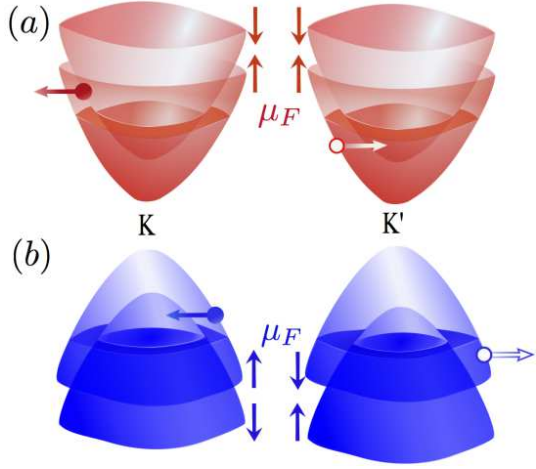}
\caption{(Color online) The band structure of spin polarized MoS$_2$-ML when the Fermi energy ($\mu_F$) lies in the conduction band (a) and in the valence band (b), respectively. During the Andreev reflection both the valley and spin degrees of the electron (denoted by filled circle) are switched in the corresponding hole (shown by empty circle) in both $n$ and $p$-doped cases. The arrows attached to the circles indicate the velocity directions of the excitaions.}\label{fig1}
\end{center}
\end{figure}

A recent experiment has shown evidences of proximity induced superconductivity in MoS$_2$ using superconducting gates at the presence of solid and liquid gating \cite{ye12}. Here, motivated by experimental and some theoretical studies \cite{taniguchi,ge, roldan, noah,majidi14,linder14,aguado11}, we investigate the Josephson effect through the ML-MoS$_2$ both at its normal and spin-polarized (magnetic) states. 
If we induce conventional s-wave superconductivity in ML-MoS$_2$ via proximity, the Cooper pairs are formed from excitations from different (opposite) valleys and having opposite spins, very similar to the case of graphene \cite{beenakker08}. This ensures that the Cooper pairs carry no momentum and net spin, in agreement with the fact that pairing occurs between time reversal partners. Subsequently, Andreev reflection which is the conversion of electron and hole excitations to each other \cite{andreev}, will switch both the valley and spin subbands (see Fig. \ref{fig1}). 
\par
In this letter, first we study the effect of doping on the critical current which reveals that for both $n$- and $p$-doped cases the supercurrent vanishes upon reaching the band edges. Nevertheless, the SOI induced intrinsic spin splitting causes differences between $n$- and $p$-dopings. The very interesting effects come to play when we apply an exchange splitting of Zeeman type which results in the simultaneous valley- and spin-polarized state for $p$-doped ML-MoS$_2$. We show that in a S$|$F$|$S Josephson junction based on ML-MoS$_2$ not only the so-called $0-\pi$ transitions \cite{buzdin} in the supercurrent can be observed, but the amplitude of supercurrent and the transition points vary with the doping. Subsequently we predict a doping induced $0-\pi$ transition which can be observed more clearly in the $p$-doped ML-MoS$_2$. 
\par
The low energy band structure of ML-MoS$_2$
consisting of the conduction and valence bands can be described by the modified Dirac Hamiltonian \cite{rostami13,kormanyos13,liu13}. In comparison to \emph{massive Dirac equation}, it contains quadratic diagonal terms beside linear off-diagonal terms in momentum ${\bf k}$ and captures the possibility of differences in the electron and hole masses. For spin $s=\pm1$ and valley labeled by $\tau=\pm1$ and 
at the presence of an exchange field $h$ the Hamiltonian reads,
\begin{eqnarray}
H_{\tau s}({\bf k})&= &\Delta\sigma_z+\lambda\tau s
(1-\sigma_z)\nonumber\\&&+\hbar v_0 {\bf k}\cdot{\bm \sigma}_\tau
+\frac{\hbar^2|{\bf k}|^2}{4m_0}(\alpha+\beta\sigma_z)-sh
\label{eq1}
\end{eqnarray}
in which the Pauli matrices ${\bm \sigma}_\tau=(\tau\sigma_x,\sigma_y)$ and $\sigma_z$ operates in a space defined by the conduction and valence bands. The band gap, the intrinsic spin-orbit coupling, and the effective velocity are given by $\Delta\sim0.95$eV, $\lambda\sim40$meV, and $v_0\sim 5\times10^5$ m/s, respectively. The electron bare mass is indicated by $m_0$ and the remaining band parameters have the values $\alpha\sim 0.43$, $\beta\sim 2.21$.
\par
From the Hamiltonian (\ref{eq1}) the eigenstates can be found as below,
\begin{eqnarray}\label{eq2}
\epsilon_{k\tau s}&=&(\lambda\tau-h) s+\frac{\hbar^2 k^2}{4m_0}\alpha \nonumber\\ &\pm&\sqrt{(\Delta-\lambda\tau s+\frac{\hbar^2k^2}{4m_0}\beta)^2+(\hbar v_0 k)^2}
\end{eqnarray}
We define the excitations energy $\epsilon>0$ with respect to the Fermi level $\mu$ so that inside the normal (or magnetic) part of ML-MoS$_2$ the excitations have the energies $\epsilon=|\mu-\epsilon_{k\tau s}|$. The band structure of ML-MoS$_2$ in the presence of an spin splitting and for both $n$ and $p$-doped cases are shown in fig. \ref{fig1}, respectively. It is clear that while for conduction band excitations we have a similar splitting of up and down spin subbands in both valleys, the valence spin subbands splitting differs in the two valleys due to the intrinsic splitting caused by SOI. This effect results in the differences between the properties of $n$ and $p$-doped ML-MoS$_2$.
\par
The Josephson junction consists of two superconducting regions 
where superconducting electrodes on top of a sheet of MoS$_2$ will induce a nonzero pairing amplitude $\Delta_{S}$ and a normal or magnetic region at the middle. The spin polarization at the middle region can be induced via proximity with a ferromagnetic insulator on top. The Josephson current is carried by the discrete bound states across the junction which are known as Andreev states. In order to obtain their energies we employ the BdG equations in the presence of $h$ to construct the scattering states in each region. The BdG equations basically describe the superconducting correlations between particle-like excitations $u$ and their corresponding time-reversed hole-like excitations $v$. For the excitations with the spin $s$ and valley $\tau$, it reads \cite{beenakker06,moghaddam08},
\begin{equation}\label{eq3}
\begin{pmatrix}
H_{\tau s}-\mu & \Delta_S(x) \\
\Delta_S^{*}(x) & \mu-\Theta H_{\tau s}\Theta^{-1}
\end{pmatrix}\begin{pmatrix}
u_{s,\tau} \\ v_{\bar{s},\bar{\tau}}
\end{pmatrix}=\epsilon\begin{pmatrix}
u_{s,\tau} \\ v_{\bar{s},\bar{\tau}}
\end{pmatrix},
\end{equation}
where $H_{\tau s}$ is the modified Dirac Hamiltonian Eq. (\ref{eq1}) and $\bar{s}=-s$, $\bar{\tau}=-\tau$. The time reversal operator $\Theta=i\tau_{x}s_{y}\mathcal{K}$ with $\mathcal{K}$ being the complex conjugate operator is brought to connect the time reversal states which has opposite spins and valleys. The two-component coherence factor of the BCS theory $u_{s,\tau}=(\psi_{c},\psi_{v})^{T}$ characterizes the particle part of the total wave function while the spinor $v_{\bar{s},\bar{\tau}}=\Theta u_{s,\tau}$ describes the corresponding hole part of it. Here $c$ and $v$ denote the conduction and valance bands, respectively. The superconducting pair potential $\Delta_S(x)$ couples $s$-particle and $\bar{s}$-hole excitations in the two valleys. As a result a spin $s$ particle from the conduction (valence) band and in valley $\tau$ will be Andreev reflected to a hole of spin $\bar{s}$ and valley $\bar{\tau}$ in the conduction (valence) 
band as it is represented clearly in Fig. \ref{fig1}. 
\par
The scattering states in the F region are as the followings,
\begin{equation}
\begin{array}{c}
\psi^{e\pm}=e^{\pm i{\bf k}_{e}\cdot{\bf r}}\left(
\eta_{e}e^{\pm i\theta_{e}}, \pm 1 , 0 , 0
\right)\\
\psi^{h\pm}= e^{\mp i{\bf k}_{h}\cdot{\bf r}}
\left(
0 , 0 , \eta_{h}e^{\pm i\theta_{h}}, \mp 1
\right)
\end{array}
\label{eq4}
\end{equation}
where $k_{e}(k_{h})$ is the electrons (holes) wave vector. The angle of incident electron and reflected hole are related via $k_{e}\sin\theta_{e}=k_{h}\sin\theta_{h}$ and we have defined $\eta_{e,h}=k_{e,h}/(\pm\epsilon+\mu_{N}\pm hs-\Delta-\hbar^{2}k_{e,h}^{2}\beta/4m_{0})$. Inside an S region the scattering basis are obtained as the followings,
\begin{equation}
\begin{array}{c}
\psi^{\tilde{e}\pm}=e^{\pm i{\bf q_{\tilde{e}}}\cdot{\bf r}}\left(
\zeta_{\tilde{e}} e^{\pm i\theta_{s}-i\Gamma},
\pm e^{-i\Gamma},\zeta_{\tilde{e}} e^{\pm i\theta_{s}+i\phi} , \pm e^{i\phi}
\right) \\
\psi^{\tilde{h}\pm}=e^{\mp i{\bf q_{\tilde{h}}}\cdot{\bf r}}\left(
\zeta_{\tilde{h}} e^{\mp i\theta_{s}+i\Gamma},\mp e^{i\Gamma},\zeta_{\tilde{h}} e^{\mp i\theta_{s}+i\phi}, \mp e^{i\phi}
\right)
\end{array}
\label{eq7}
\end{equation}
where $q$ is the wave vector and $\tilde{e}$, $\tilde{h}$ represent the  electron- and the hole-like Bogoliubov quasiparticles, respectively. We note also their angles is related to the angle of incidence via conservation of the transverse component of the wave vectors by the translation invariance. $\Gamma=\cos^{-1}(\epsilon/\Delta_{S0})$ and $\phi$ denote the Andreev and superconducting phases, respectively and for the sake of clarity we define $\zeta_{\tilde{e},\tilde{h}}=(\mu_{s}+\Delta-2\lambda s+\hbar^{2}q_{\tilde{e},\tilde{h}}^{2}\beta/4m_{0}\pm\sqrt{\epsilon^{2}-\Delta_{S0}^{2}})/q_{\tilde{e},\tilde{h}}$.
\par
\begin{figure}[tp]
\begin{center}
\includegraphics[width=0.8\linewidth]{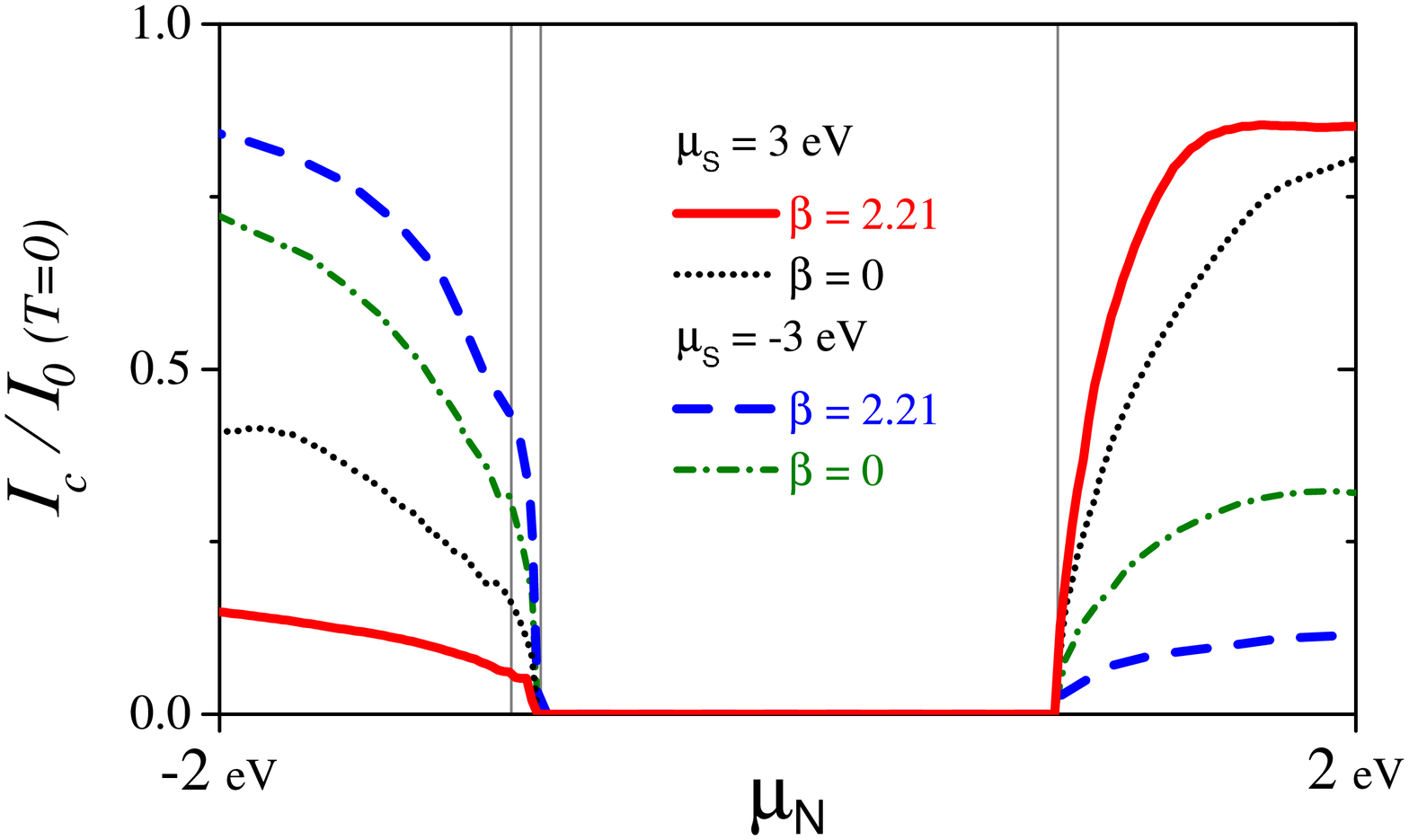}
\caption{(Color online) Critical current $I_{c}/I_{0}(T)$ versus chemical potential $\mu_{N}$ is depicted in zero temperature for a junction of length $L/\xi=0.05$. Two superconductor either $n$-doped or $p$-doped are connected by a short junction which made of a $n$-doped normal layer (black and red curves), and by a $p$-doped layer (blue and green curves). The gray lines not only determine the edge of the conduction and two spin split valance bands but represent also the band gap in ML-MoS$_2$.}\label{fig2}
\end{center}
\end{figure}
\par
We are now in a position to write down the wavefunctions in each region as a linear superposition of constructed scattering states Eqs. (\ref{eq4}) and (\ref{eq7}),
\begin{eqnarray}
\psi_{L}&=&t_{1}\psi_{S}^{\tilde{e}-}+t_{2}\psi_{S}^{\tilde{h}-}\nonumber\\
\psi_{F}&=&t_{3}\psi^{e+}+t_{4}\psi^{h+}+t_{5}\psi^{e-}+t_{6}\psi^{h-}\nonumber\\
\psi_{R}&=&t_{7}\psi_{S}^{\tilde{e}+}+t_{8}\psi_{S}^{\tilde{h}+}
\end{eqnarray}
The Andreev bound states may now be computed by evaluating the determinant of the coefficient matrix by matching the wavefunctions at the interfaces $x=0$ and $x=L$ (as follows from conservation of current flux) and subsequently used to find the carried current across the junction in the presence of an exchange field.
\par
In the short junction limit $L\ll \xi$, where $\xi\sim\hbar v_0/\Delta_{S0}$ is the superconducting coherence length, the main contribution to the supercurrent comes from the discrete spectrum corresponding to bound states in the gap. The current $I$ passing through the junction with transverse width $W$ is given by,
\begin{equation}\label{current}
I=-\frac{2eW}{\hbar}\int dk_{y}\sum_{n s}\tanh[\frac{\epsilon_{n s}(k_{y})}{2k_{B}T}]\frac{d\epsilon_{n s}(k_{y})}{d\varphi}
\end{equation}
where $\epsilon_{n s}$ is the Andreev bound state for spin spices-$s$ and $n$ denote the mode number while $T$ is the temperature. As usual the supercurrent is derived by the superconducting phase difference $\varphi=\phi_{L}-\phi_{R}$. The sum is over all states with positive eigenvalues corresponding to the different transverse momentum $k_y$. We scale the supercurrent with $I_{0}(T)=\frac{e \Delta(T)}{\hbar}W(k_{F\uparrow}+k_{F\downarrow}) $ in which $k_{F s}$ indicates the Fermi wave vector of spin $s$ carriers at the non-magnetic state $h=0$.
\par
First we study the supercurrent passing through ML-MoS$_2$ without any exchange splitting. We remind that in all of the numerical calculations we drop the $\alpha$-term in the Hamiltonian since it does not play an important role in the transport properties. 
\par
Figure \ref{fig2} shows the variation of critical current $I_c$ versus the chemical potential of normal region ($\mu_N$) at zero temperature. As we expect inside the gap region $-\Delta+\lambda<\mu_N<\Delta$ the transport of Cooper pairs is completely suppressed. However for the heavily doped case in which the Fermi level lies inside the conduction or valence bands a finite supercurrent can pass through ML-MoS$_2$, which shows a fast decline upon reaching the band edges. Moreover we see an asymmetry between electron and hole doping cases so that $I_c$ is larger when the normal region has the same type of carriers as the superconducting regions. In other words when the carriers type between N and S regions differs the particles pass through a $npn$ or $pnp$ junction which are less conducting. But intriguingly the supercurrent in these cases is more influenced  by the inclusion of topological term $\beta$ and in addition it is not suppressed significantly unlike conventional $npn$ and $pnp$ junctions. All these behaviors are the characteristics of modified Dirac Hamiltonian which governs the excitations in ML-MoS$_2$.
\par
\begin{figure}[tp]
\begin{center}
\includegraphics[width=0.8\linewidth]{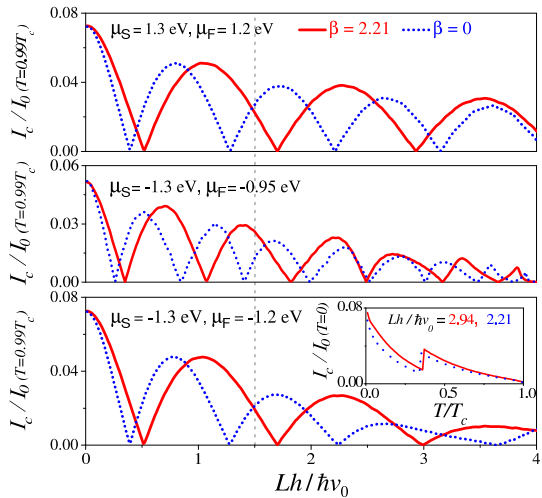}
\end{center}
\caption{(Color online) Critical current $I_{c}/I_{0}(T)$ versus dimensionless length $Lh/\hbar v_0$ is depicted in the near transition temperature limit $T=0.99T_{c}$. Two superconductor either $n$-doped or $p$-doped are connected by a short junction which made of a $n$-doped ferromagnetic layer (top panel), a $p$-doped ferromagnetic layer (middle panel), a $p$-doped ferromagnetic layer (bottom panel). Critical current goes to zero at transition point between $0$ and $\pi$ phases. The Inset shows the variation of the critical current with temperature revealing the possiblity of temperature induced $0-\pi$ transition.}\label{fig3}
\end{figure}

When we induce an exchange splitting in the middle region the supercurrent can change the sign and a $0-\pi$ transitions occurs. Varying this splitting or the length of the junction successive $0-\pi$ transitions occur and the critical current shows an oscillatory behavior as a function of $Lh/\hbar v_0$. Such behavior is shown in Fig. \ref{fig3} where $I_c$ is plotted as a function of $Lh/\hbar v_0$ for $T=0.99T_c$. It is well known that in the $0-\pi$ transition points only the contribution coming from the first harmonic $\sin \varphi$ changes its sign. 
As a result, since close to the critical temperatures $T=0.99T_c$ the Josephson current shows a sinusoidal phase dependence, the critical current vanishes at the transition points. 
\par
In order to see the effect of doping on the critical current of SFS system, the critical current is found for different situations in which the system is $n$- and $p$-doped, respectively. The main difference between them is in the positions of the transition points. In addition the effect of $\beta$-term is also to shift the transitions points in both the $n$- and $p$-doped systems. In the three different situations of Fig. \ref{fig3} the Fermi level of the ferromagnetic region crosses, respectively, the conduction band ($\mu_F=1.2$ eV), one of the spin-split valence subbands ($\mu_F=-0.95$ eV), and finally both of the valence subbands ($\mu_F=-1.2$ eV). Although the highly doped cases ($\mu_F=\pm1.2$ eV) behave more or less the same, when the Fermi level lies between the spin-split valence subbands, the critical current is weaker and its oscillations have a smaller period. The decline in the amplitude of $I_c$ can be understood from the fact that the density of states in this case is smaller by a factor of almost two since only one subband participates in the transport effectively. 
\par
The dependence of oscillatory variations on the doping is one of our key findings. In fact it shows that by varying only the Fermi level of the magnetic region by some gate voltages for instance, $0-\pi$ transitions can occur. For example at $Lh/\hbar v_0=1.5$, as indicated by a vertical dashed line, the three dopings in Fig. \ref{fig3} correspond to $\pi$, $0$, and $\pi$ states, respectively.
But such a behavior can be understood noting to the fact that the period of oscillations versus length which is called magnetic coherence length $\xi_h\sim\hbar v_{\rm F}/h$, varies with the Fermi velocity $v_{\rm F}=(1/\hbar)\partial \epsilon_k/\partial k$. Invoking the dispersion relation (\ref{eq2}) we find the following form for the magnetic coherence length as a function of Fermi wave vector,
\begin{eqnarray}\label{xih}
\xi_{h}=\frac{1}{2}&&\sum_{s=\uparrow,\downarrow}\frac{\hbar^2 k_{F s}}{2m_{0} h}\nonumber\\
&&\left\vert\alpha\pm\frac{(\Delta-\lambda s+\frac{\hbar^2k_{F s}^2}{4m_0}\beta) \beta+ 2 m_0 v_{0}^{2}  }{\sqrt{(\Delta-\lambda s+\frac{\hbar^2k_{F s}^2}{4m_0}\beta)^2+(\hbar v_0 k_{F s})^2}}\right\vert.~~~~
\end{eqnarray}
The Fermi velocity at different chemical potentials used in our calculations \emph{i.e.} $\mu_{N}=1.2, -0.95, -1.2$eV, is found $v_{F}/v_{0}\simeq 0.81, 0.25, 0.8$ for $\beta=2.21$ and $v_{F}/v_{0}\simeq 0.61, 0.2, 0.6$ for $\beta=0$, respectively. This clearly shows the period of oscillations $\xi_h$ for intermediate doping $\mu=-0.95$ is significantly smaller and in addition the $\beta$-term increases the period in all cases. In contrast, the $0-\pi$ transition points very weakly depend on the doping strength of superconducting regions. It is worthnoting that the relation (\ref{xih}) indicates that the $\alpha$-term only changes the $\xi_h$ slightly. As indicated in the the inset of Fig. \ref{fig3} by an abrupt change in $I_c$ we can have the temperature induced transition, too. This originates from the fact that the transition points versus $Lh/\hbar v_0$ show a weak dependence on the temperature. Beside this in general the critical current decreases with $T$ and vanishes at $T=T_c$ as it is expected. 
\par
Finally we comment on the relevance of results obtained here to other 2D Dirac systems. In the case of MoS$_2$ there are clear evidence of superconducting correlations induction which increases the possibility of experimental verification of our results. However due to their layered nature, such correlations can be in principle induced in other 2D systems like recently synthesized silicene, germanene \cite{ni12} and phosphorene \cite{liu14} as well as other transition metal dichalcogenides. So the Josephson coupling and also $0-\pi$ transitions with length variation at the presence of magnetic correlations can be observed in them. But the doping induced transitions may be difficult to be seen in above mentioned 2D materials, although further investigations are needed to find the behavior of supercurrent in them. In fact as discussed above the significant change in the magnetic coherence length or in other words the effective Fermi velocity is needed for such transitions. In MoS$_2$ due to the spin-splitting of valence band $\xi_h$ decreases significantly when the Fermi level crosses only one subband (intermediate hole-doping) as one can see from faster oscillations in the middle panel of Fig. \ref{fig3}. Therefore since large spin splitting of valence band is almost a unique feature of MoS$_2$ monolayer it is more likely to observe doping induced $0-\pi$ transition in it rather than other 2D Dirac systems. 
\par 
In conclusion the Josephson effect in the monolayer MoS$_2$ junctions in the presence and absence of extrinsic spin splitting is studied. The critical current is found to be slightly different for $n$ and $p$ doped cases mostly due to the 
intrinsic spin splitting in the valence band originating from spin-orbit interaction. When we apply the external Zeeman splitting $h$ via proximity for instance, the successive $0-\pi$ transitions are found indicated by an oscillatory behavior for the critical current as a function of $Lh/\hbar v_0$. We see that both the amplitude and the period of oscillations vary with the doping due to the change in the Fermi velocity. This provides the possibility of observing $0-\pi$ transitions by changing only the doping strength, beside the temperature induced transition.


\end{document}